\pdfoutput=1
\documentclass[journal]{IEEEtran}

\usepackage{amsmath}
\usepackage{amssymb}
\usepackage{graphicx}
\usepackage{booktabs}
\usepackage{multirow}
\usepackage{cite}
\usepackage{url}

\usepackage{hyperref}
\usepackage{array}
\usepackage{textcomp}
\usepackage{float}
\usepackage{tabularx}
\usepackage{xcolor}
\usepackage{balance}
\usepackage{makecell}
\usepackage{longtable}

\hypersetup{
    colorlinks=true,
    linkcolor=black,
    citecolor=black,
    urlcolor=blue,
    pdfauthor={Andrii Vakhnovskyi},
    pdftitle={AI-Driven CEA as Resilient Infrastructure for U.S. Fresh-Produce Supply Chains}
}

\newcommand{\etal}{\textit{et~al.}}

\begin{document}

% ---- Title ----
\title{AI-Driven Controlled Environment Agriculture as Resilient Infrastructure for U.S.\ Fresh-Produce Supply Chains}

% ---- Author ----
\author{ANDRII VAKHNOVSKYI%
\thanks{A.~Vakhnovskyi is with IOGRU LLC, New York, NY 10022, USA (e-mail: andrii.vakhnovskyi@gmail.com). ORCID: 0009-0007-8306-5932.}%
}

\maketitle

% ===========================================================================
% ABSTRACT
% ===========================================================================
\begin{abstract}
Climate volatility, regional production concentration, labor constraints, cyber risk, and dependence on long-distance fresh-produce supply chains expose vulnerabilities in U.S.\ fresh-produce and specialty-crop systems. Controlled environment agriculture (CEA) can reduce some exposure to weather and logistics disruptions by moving selected production into protected, sensor-rich environments, but recent failures in venture-backed vertical farming show that CEA cannot be treated as a universal or automatically sustainable food-security solution. This paper proposes the Controlled Environment Agriculture Resilience Infrastructure Framework, Version~2.0 (CEA-RIF~2.0), for evaluating AI-driven CEA as targeted regional fresh-produce continuity infrastructure. The framework assesses seven dimensions: supply continuity, climate isolation, energy and grid integration, water and nutrient circularity, cyber-physical reliability, economic viability, and governance and deployment. Drawing on U.S.\ government reports, peer-reviewed CEA and energy literature, demand-response research, cybersecurity standards, international smart-agriculture programs, 2025--2026 financing and policy signals, and public autonomous-greenhouse datasets, the paper argues that AI creates resilience value only when it improves measured operational outcomes such as climate stability, energy flexibility, yield consistency, anomaly detection, labor productivity, and safe recovery from faults. The analysis reframes AI-driven CEA as a cyber-physical infrastructure problem: energy-aware, grid-interactive, secure, interoperable, regionally distributed, financially disciplined, and connected to public resilience goals. The paper concludes with a research agenda for interagency testbeds, open datasets, standardized metrics, demand-response pilots, cyber-physical reference architectures, and public-private deployment models that can distinguish viable resilience assets from fragile business models.
\end{abstract}

% ---- Index Terms ----

% ===========================================================================
% SECTION 1: INTRODUCTION
% ===========================================================================
\section{Introduction}

\IEEEPARstart{U}{.S.} food-system resilience is often discussed through the lens of staple commodities, aggregate calories, or national production totals. Those frames are important, but they can obscure a different vulnerability: fresh produce and specialty crops are perishable, climate-sensitive, labor-intensive, and dependent on continuous logistics. Leafy greens, tomatoes, cucumbers, peppers, berries, herbs, seedlings, and propagation material do not behave like shelf-stable commodities. They depend on cold chains, skilled labor, water availability, plant health, regional harvest windows, and reliable transportation. When drought, heat, wildfire smoke, floods, disease outbreaks, port delays, border disruptions, labor shortages, cyber incidents, or extreme weather affect production and distribution, communities can lose access to fresh, nutritious, and institutionally important foods even when total national calorie availability remains adequate.

Controlled environment agriculture (CEA) offers one response to this problem, but only if its value is defined carefully. CEA includes greenhouses, indoor farms, plant factories, vertical farms, hydroponic systems, aeroponic systems, aquaponic systems, and related protected-production formats in which environmental variables such as light, temperature, humidity, carbon dioxide, air movement, water, and nutrients are partially or fully managed through physical infrastructure and control systems. These systems can reduce exposure to some outdoor climate risks, extend growing seasons, enable production near demand centers, and improve control over water and nutrients. At the same time, CEA introduces new dependencies: electricity, HVAC, lighting, sensors, controllers, cloud services, vendor platforms, cybersecurity, skilled operators, and capital-intensive facility economics.

This paper therefore does not argue that CEA or vertical farming will replace conventional agriculture. It does not argue that CEA can solve U.S.\ food security, eliminate imports, or make food systems automatically sustainable. The stronger and more defensible claim is narrower: AI-driven CEA can become a targeted resilience layer for selected U.S.\ fresh-produce and specialty-crop supply chains if it is designed and evaluated as cyber-physical infrastructure. That means energy-aware, grid-interactive, economically disciplined, secure, interoperable, regionally distributed, and connected to public resilience goals.

The need for discipline is urgent because CEA sits at the intersection of real promise and real failure. U.S.\ government sources recognize Food and Agriculture as a critical infrastructure sector, with dependencies on water, transportation, energy, chemicals, and information technology~\cite{ref_fda_sector}. USDA and DOE have identified indoor agriculture as an interdisciplinary research area~\cite{ref_usda_doe_workshop}, and recent CEA research emphasizes interagency collaboration across agriculture, energy, space research, food safety, standards, and workforce development~\cite{ref_boyd_pnas}. At the same time, several prominent vertical-farming companies have filed for Chapter~11, restructured, shut down, or retrenched~\cite{ref_appharvest,ref_kalera,ref_aerofarms,ref_bowery,ref_plenty,ref_infarm,ref_fifth_season,ref_upward,ref_smallhold,ref_vertical_future}. These failures do not invalidate protected cultivation, greenhouse agriculture, propagation, indoor plant science, or AI-assisted CEA control. They do show that any infrastructure argument must be built on measurable performance rather than hype.

This paper contributes CEA-RIF~2.0, a resilience infrastructure framework for evaluating when AI-driven CEA creates public and regional value. The framework has seven dimensions: supply continuity, climate isolation, energy and grid integration, water and nutrient circularity, cyber-physical reliability, economic viability, and governance and deployment. It also introduces two deployment concepts. The first is \textit{grid-aware CEA}: CEA designed to coordinate crop performance with electricity-system constraints through forecasting, load shifting, demand response, energy storage, renewable integration, and control policies that protect plant health. The second is the \textit{regional fresh-produce continuity node}: a CEA facility or network designed to maintain selected fresh produce, seedlings, or specialty crops during normal operations and under defined disruption scenarios while meeting transparent energy, water, economic, cyber-physical, and governance standards.

The paper is organized as follows. Section~II discusses CEA after the hype cycle, including the 2025--2026 market and financing stress test. Section~III defines the U.S.\ fresh-produce and specialty-crop resilience problem. Section~IV summarizes international lessons. Section~V describes the method as a structured evidence synthesis and framework-development process. Section~VI presents CEA-RIF~2.0. Sections~VII through~X examine AI functions, grid-aware CEA, cyber-physical reliability, and regional continuity nodes. Section~XI provides supporting tables for evaluation and governance. Sections~XII through~XVI discuss limitations, research priorities, data and code availability, AI assistance disclosure, and conclusions.

% ===========================================================================
% SECTION 2: CEA AFTER THE HYPE CYCLE
% ===========================================================================
\section{Controlled Environment Agriculture After the Hype Cycle}

The strongest case for CEA must begin by acknowledging failure. Since 2022, at least ten venture-backed vertical-farming and indoor-agriculture companies have entered bankruptcy, restructuring, shutdown, or liquidation across the United States and Europe. AppHarvest filed for Chapter~11 in July~2023 and was liquidated, with facilities sold to Equilibrium, Bosch Growers, and Mastronardi Produce~\cite{ref_appharvest}. Kalera's main operating subsidiary filed for Chapter~11 in April~2023; its assets were acquired by Sandton Capital Partners and later resold to 80~Acres Farms in April~2025~\cite{ref_kalera}. AeroFarms filed for Chapter~11 in June~2023 and emerged in September~2023, but a December~2025 WARN notice reported that its Virginia operating entities planned to wind down operations after a largest-investor funding withdrawal; the company subsequently reported short-term stakeholder funding to continue operations and explore strategic alternatives~\cite{ref_aerofarms}. Bowery Farming, which had raised over \$700~million at a peak valuation of \$2.3~billion, ceased all operations in November~2024 without filing for bankruptcy~\cite{ref_bowery}. Plenty Unlimited filed a prepackaged Chapter~11 in March~2025 and emerged in May~2025, narrowing its focus to premium strawberry production~\cite{ref_plenty}. Infarm declared insolvency across multiple European jurisdictions in 2023; UK Companies House records show the original foreign company administration ran from September~2023 to September~2025, while successor Infarm Technologies Limited became subject to compulsory liquidation in February~2025~\cite{ref_infarm}. Fifth Season shut down in October~2022 and filed Chapter~11 in November~2023~\cite{ref_fifth_season}. Upward Farms ceased operations in April~2023~\cite{ref_upward}. Smallhold filed Chapter~11 in February~2024~\cite{ref_smallhold}. Vertical Future entered administration in August~2025 after revenue collapsed from GBP~6.7~million to GBP~692,000, and entered creditors' voluntary liquidation in January~2026~\cite{ref_vertical_future}.

Not all CEA operators have failed. Local Bounti (NASDAQ: LOCL) restructured its debt in 2024 and reported \$48.4~million in full-year 2025 revenue, a 27\% increase, while also disclosing continuing risks typical of a capital-intensive early-growth company~\cite{ref_local_bounti}. Gotham Greens operates 13 hydroponic greenhouse facilities totaling over 1.8~million square feet. 80~Acres Farms acquired three former Kalera vertical farms and related intellectual property in 2025, and later announced a strategic merger with Soli Organic that combined automated production systems with an established retail footprint and decades of commercial production experience~\cite{ref_80acres}. These events suggest a consolidation trend in which operators with stronger commercial channels, standardized technology, and agronomic execution may absorb or repurpose assets from failed ventures. The lesson is not that CEA is dead; it is that undisciplined business models fail while operationally sound models may survive.

These events should not be used as a simplistic argument that CEA is dead. Greenhouse agriculture is a mature sector in parts of North America, Europe, and Asia. Protected cultivation ranges from low-cost plastic structures to highly automated glasshouses. Indoor plant factories may still be useful for research, propagation, high-value products, harsh environments, or tightly controlled production. The lesson is not that CEA lacks technical value; it is that CEA must be evaluated as infrastructure and operations, not as a growth story detached from energy, crop economics, labor, logistics, biological risk, and control reliability.

Post-hype CEA requires a viability filter. A facility should not be considered a resilience asset merely because it grows indoors or uses AI. It should be evaluated against transparent metrics for energy intensity, load flexibility, crop economics, water use, labor productivity, uptime, cyber maturity, supply continuity, and governance. Resilience value must be earned through measurement.

Recent federal financing and insurance signals sharpen this point. In April~2026, USDA Rural Development extended its pause on accepting, processing, and awarding Rural Business-Cooperative Service loan note guarantees for biodigester and CEA projects through December~31, 2026, citing elevated portfolio risk and reporting that a recent snapshot showed a 40\% delinquency rate for controlled-environment agriculture projects~\cite{ref_usda_rd_pause}. That action should not be read as a rejection of all CEA. It should be read as a warning that credit underwriting must account for engineering completeness, anchor-buyer commitments, repayment capacity, asset standardization, and the limited secondary market for bespoke facility equipment. At the same time, USDA's Risk Management Agency expanded the Controlled Environment pilot crop insurance program to an additional 48 counties in 17 states for the 2026 and succeeding crop years, increased the upper coverage limit from 75\% to 85\%, and included coverage for plant diseases subject to destruction orders~\cite{ref_usda_rma}. The combined signal is mixed but useful: CEA is becoming more legible to federal agricultural programs, while poorly underwritten projects are being treated as high risk. CEA-RIF~2.0 should therefore evaluate both insurability and bankability, not only technical performance.

% ===========================================================================
% SECTION 3: THE RESILIENCE PROBLEM
% ===========================================================================
\section{The U.S.\ Fresh-Produce and Specialty-Crop Resilience Problem}

The U.S.\ Food and Agriculture Sector is recognized as one of sixteen critical infrastructure sectors under the national framework established by Presidential Policy Directive~21 (2013) and updated by National Security Memorandum~22 (April~2024)~\cite{ref_fda_sector,ref_nsm22}. FDA describes the sector as including farms, manufacturers, processors, storage and warehousing facilities, restaurants, retail establishments, and more, with critical dependencies on water, transportation, energy, chemicals, and information technology. USDA and HHS/FDA serve as co-Sector Risk Management Agencies. In July~2025, USDA Secretary Brooke Rollins launched the National Farm Security Action Plan, declaring that ``farm security is national security''~\cite{ref_usda_farm_security}, and in February~2026, DARPA and USDA signed a memorandum of understanding establishing collaboration on agriculture-related research and development initiatives to safeguard and defend the U.S.\ food system~\cite{ref_darpa_usda}. This is a broad sector, and CEA is not itself a separate critical infrastructure sector. The correct framing is that CEA is an emerging production modality within a sector that the federal government now treats as both critical infrastructure and a national security priority.

Fresh produce is especially relevant because it is perishable and disruption-sensitive. USDA ERS reported that from 2007 to 2023, the import share of U.S.\ fresh fruit availability grew from 50\% to 59\% and the import share of fresh vegetables, excluding potatoes, sweet potatoes, and mushrooms, grew from 20\% to 35\%~\cite{ref_usda_ers_imports}. In 2023, Mexico and Canada supplied most fresh vegetable imports by value. Domestic production is also geographically concentrated. Approximately 90\% of U.S.\ lettuce comes from two regions: the Salinas Valley in California during spring and summer and the Yuma area of Arizona during winter, with the same equipment, workers, and processing plants physically migrating between seasons~\cite{ref_usda_ers_eib264}. This concentration has contributed to at least six distinct multi-state \textit{E.\ coli} O157:H7 outbreaks linked to romaine lettuce between 2017 and 2022, including a 2018 Yuma-linked outbreak that caused 234 illnesses across 33 states. The COVID-19 pandemic disrupted fresh-produce supply chains through labor shortages, foodservice-channel collapse, and cold-chain bottlenecks, prompting USDA to deploy \$19~billion in emergency food assistance. Biological shocks also expose food-system fragility. The ongoing HPAI avian influenza outbreak, the largest in U.S.\ history, had affected more than 179~million birds by September~30, 2025, according to USDA APHIS~\cite{ref_usda_aphis_hpai}, and EFSA reported 2,896 HPAI A(H5) detections in domestic and wild birds across 29 European countries between September~6 and November~28, 2025~\cite{ref_efsa_hpai}. HPAI primarily affects poultry and should not be used to imply that CEA solves animal-disease risk. It is relevant here because it illustrates how vector pathways, environmental exposure, and regional biosecurity failures can destabilize food infrastructure. For plant production, enclosed or semi-enclosed CEA can reduce some exposure pathways through filtration, sanitation, water-quality control, pest exclusion, and compartmentalization, but it cannot guarantee pathogen-free operation.

Climate stress increases exposure. USDA Climate Hubs describes specialty crops as ``generally more sensitive to climatic stressors and more management-intensive than traditional row crops''~\cite{ref_usda_climate_hubs}. Kistner~\etal\ found that temperature and precipitation fluctuations directly affect specialty-crop production quantity and quality~\cite{ref_kistner}.

Cybersecurity threats to food and agriculture are also escalating. The JBS Foods ransomware attack in 2021 shut down 13~U.S.\ meat plants and resulted in an \$11~million ransom payment. The bipartisan Farm and Food Cyber\-security Act of 2025 (S.754/H.R.1604) would direct USDA to conduct biennial cyber\-security threat assessments and annual cross-sector crisis simulation exercises~\cite{ref_farm_cyber_act}.

CEA can reduce some exposure by creating controlled production capacity for selected crops, particularly where regional continuity matters for hospitals, schools, food banks, military bases, remote communities, disaster-prone regions, or vulnerable populations. CEA also introduces dependencies that can become resilience weaknesses. Fully enclosed vertical farms can be electricity intensive because artificial lighting replaces sunlight and HVAC/de\-hu\-mid\-i\-fi\-ca\-tion must remove heat and moisture. Cyber-physical dependence grows as facilities add sensors, remote dashboards, AI models, and vendor access. The resilience question is therefore not whether CEA is protected from the outdoor environment. The question is whether the facility improves regional supply continuity after accounting for energy, economics, cyber-physical reliability, and governance.

% ===========================================================================
% SECTION 4: INTERNATIONAL LESSONS
% ===========================================================================
\section{International Lessons}

International evidence supports the idea that AI-enabled CEA and smart agriculture are strategic, but it also reinforces the need for systems thinking. The Netherlands emphasizes mature greenhouse horticulture, energy transition, and autonomous greenhouse benchmarking. Singapore links high-technology local production to food security under import dependence, land scarcity, and resource constraints~\cite{ref_singapore_sfa}. Gulf states connect modern production technologies to arid climate and supply resilience. Japan and South Korea emphasize smart agriculture, robotics, data platforms, and workforce renewal. Canada demonstrates that controlled-environment vegetable production is already a serious North American industry. Australia frames protected cropping around climate resilience. The EU contributes a governance and interoperability lens through agricultural digitalization and data-space initiatives.

The common lesson is that successful CEA is not just a farm technology. It depends on energy systems, workforce pipelines, public-private research, market channels, data governance, standards, financing, and regional planning. The United States does not need to copy Singapore, the Netherlands, Japan, Korea, or Canada. It needs a U.S.-specific framework for deciding where AI-driven CEA provides measurable resilience value under American risks: hurricanes, drought, heat, wildfire smoke, port disruptions, transportation bottlenecks, cyber incidents, skilled labor shortages, and energy constraints.

Table~\ref{tab:international} summarizes international lessons and transfer risks.

\begin{table*}[!t]
\centering
\caption{International Lessons for U.S.\ Deployment}
\label{tab:international}
\footnotesize
\begin{tabular}{@{}p{1.8cm}p{2.8cm}p{3.8cm}p{3.8cm}p{3.2cm}@{}}
\toprule
Country/Region & Strategic Driver & CEA Mechanism & U.S.\ Lesson & Transfer Risk \\
\midrule
Netherlands & Greenhouse expertise, energy transition & Autonomous greenhouse research, high-tech glasshouses, energy-aware programs & Treat AI control as a systems problem involving plant physiology, climate models, grower expertise, and energy & Dutch greenhouse economics do not directly transfer to all U.S.\ regions \\
Singapore & Import dependence ($>$90\%), land scarcity & High-tech local production; 2035 targets of 20\% fibre, 30\% protein & CEA is part of food-security portfolio but faces real cost and scale constraints & City-state conditions differ from U.S.\ regional systems \\
UAE/Qatar & Arid climate, water scarcity, supply resilience & Modern agriculture, hydroponics, vertical farming, strategic reserves & CEA can support resilience in climate-extreme settings & Public claims can be promotional \\
Japan/Korea & Aging workforce, labor shortage & Smart Farm Innovation Valleys, robots, AI, data platforms & AI can preserve and augment scarce grower expertise & Farm structure differs from U.S.\ specialty-crop systems \\
Canada & Year-round produce, greenhouse exports & Greenhouse vegetable industry, automation & CEA is already a North American competitiveness issue & Cluster-specific market and energy conditions \\
EU & Sustainability, digital sovereignty & Agricultural digitalization, common data spaces, interoperability & Regional CEA requires trusted data sharing and standards & EU data governance may not map to U.S.\ private-sector systems \\
Australia & Climate variability, energy efficiency & Protected-cropping research hubs, AgriPV, smart films & Energy-smart protected cropping should be central to research design & Different climate zones and policy institutions \\
China & Scale, rural modernization, food-system modernization & Large greenhouse expansion, smart-agriculture action plans, AI/big-data/IoT policy & Global competition increases the need for secure, interoperable, energy-aware U.S.\ systems & Do not equate all greenhouse area with high-tech AI CEA \\
\bottomrule
\end{tabular}
\end{table*}

% ===========================================================================
% SECTION 5: METHODS
% ===========================================================================
\section{Methods: Structured Evidence Synthesis and Framework Development}

This paper is designed as a framework paper supported by structured evidence synthesis. The method has five steps.

First, sources are grouped into six evidence streams: U.S.\ food and agriculture critical infrastructure policy; CEA trends and economics; energy and grid-interactive indoor agriculture; AI, automation, and autonomous greenhouse control; cybersecurity and OT standards; and international smart-agriculture policy.

Second, each evidence stream is mapped to recurring constraints: energy intensity, carbon intensity, crop economics, labor, water, nutrient management, data quality, interoperability, cybersecurity, vendor dependence, regional demand, and governance.

Third, the constraints are translated into CEA-RIF~2.0 dimensions. A dimension is included only if it affects whether AI-driven CEA can provide resilience value beyond normal private production.

Fourth, the framework is applied to scenario archetypes: an urban leafy-greens node near institutional customers; a greenhouse tomato or cucumber cluster with demand-response participation; a seedling and transplant node for post-disaster recovery; and a remote or climate-vulnerable community production node.

Fifth, the paper identifies metrics that can be measured in future empirical work. The present version combines a structured review, standards map, framework, scenario analysis, and demonstrative open-data metric using the Wageningen/4TU Autonomous Greenhouse Challenge dataset~\cite{ref_wur_4tu_2nd}. Future versions could add de-identified deployment data only if permission exists and confidentiality can be protected.

\subsection{Demonstrative Open-Data Metric}

To test whether a publishable mini-analysis is practical, this paper uses the 4TU/WUR Autonomous Greenhouse Challenge, Second Edition (2019) dataset rather than the much larger third-edition lettuce dataset. The second-edition dataset is small enough for rapid analysis and includes outdoor and indoor greenhouse climate, irrigation, actuator status, requested and realized climate setpoints, resource consumption, harvest, crop parameters, tomato quality, irrigation/drain samples, and root-zone information. The experiment compared five AI-controlled greenhouse compartments with a reference compartment operated by Dutch commercial growers during a cherry tomato crop.

As a demonstration only, vapor pressure deficit (VPD) was calculated from five-minute air temperature and relative humidity data using:
\begin{equation}
\text{VPD} = \left(1 - \frac{\text{RH}}{100}\right) \cdot 0.6108 \cdot e^{\frac{17.27 \, T_a}{T_a + 237.3}}
\label{eq:vpd}
\end{equation}

Table~\ref{tab:vpd_metrics} reports an illustrative VPD stability metric: the percentage of samples within a 0.5--1.2~kPa band. This band is not proposed as a final tomato-production standard; it is used to show how CEA-RIF can translate raw greenhouse climate data into a transparent resilience metric. Future empirical extensions should replace this illustrative band with crop-stage-specific thresholds from horticultural literature or expert review.

\begin{table}[!t]
\centering
\caption{Illustrative VPD Stability and Resource Use (WUR/4TU 2nd Edition)}
\label{tab:vpd_metrics}
\scriptsize
\setlength{\tabcolsep}{3pt}
\begin{tabular}{@{}lrrrrrr@{}}
\toprule
Compart. & $\bar{T}$ & $\overline{\text{RH}}$ & $\overline{\text{VPD}}$ & VPD$_{0.5\text{-}1.2}$ & Heat & Light \\
 & (\textdegree{}C) & (\%) & (kPa) & (\%) & (MJ/m$^2$) & (kWh/m$^2$) \\
\midrule
AICU & 22.06 & 84.17 & 0.45 & 26.46 & 252.30 & 240.26 \\
Automatoes & 23.27 & 84.93 & 0.47 & 25.67 & 185.31 & 270.45 \\
Digilog & 21.39 & 81.34 & 0.50 & 28.94 & 172.99 & 323.16 \\
IUACAAS & 21.32 & 84.18 & 0.45 & 29.23 & 334.82 & 228.03 \\
Reference & 22.71 & 81.76 & 0.54 & 28.78 & 471.56 & 267.20 \\
Automators & 21.35 & 80.76 & 0.53 & 29.82 & 362.60 & 284.80 \\
\bottomrule
\end{tabular}
\end{table}

The initial result is not a claim that one team was more resilient than another. It shows that open CEA datasets can support reproducible metrics linking climate stability, resource consumption, and control strategies. A subsequent empirical study should define crop-stage-specific VPD and temperature bands, normalize energy and water metrics by yield or production area, and report uncertainty.

% ===========================================================================
% SECTION 6: CEA-RIF 2.0 FRAMEWORK
% ===========================================================================
\section{CEA-RIF 2.0 Framework}

CEA-RIF~2.0 evaluates whether AI-driven CEA functions as resilience infrastructure. It does not ask whether a facility is technologically impressive. It asks whether the system provides measurable, governable, economically credible resilience value. The seven dimensions are presented in Table~\ref{tab:cea_rif}.

\begin{table*}[!t]
\centering
\caption{CEA-RIF 2.0 Framework}
\label{tab:cea_rif}
\footnotesize
\begin{tabular}{@{}p{2cm}p{2.8cm}p{2.8cm}p{2.5cm}p{2.5cm}p{2.5cm}@{}}
\toprule
Dimension & Core Question & Example Metrics & AI Contribution & Failure Mode & Governance Req. \\
\midrule
Supply continuity & Can facility maintain supply under disruptions? & kg/m$^2$; days of local supply; recovery time; contract coverage & Yield forecasting; scheduling; demand matching & Narrow crop with little resilience value & Define disruption scenarios and service-level targets \\
Climate isolation & Does system reduce exposure to heat, drought, smoke, pests? & VPD stability; pest incidents; crop-cycle completion rate & Predictive control; disease detection; anomaly detection & Indoor instability replaces outdoor risk & Document target bands and biosecurity pathways \\
Energy \& grid integration & Can production occur at acceptable cost with flexible load? & kWh/kg; peak kW; DR capacity; carbon-weighted energy & MPC; price-aware scheduling; HVAC optimization & Electricity cost undermines resilience claim & Require transparent energy/carbon accounting \\
Water \& nutrient circularity & Does system improve water/nutrient performance? & L/kg; recirculation rate; NUE; discharge volume & Irrigation optimization; leak detection; recipe control & Recirculation spreads pathogens & Require water-quality monitoring \\
Cyber-physical reliability & Can system remain safe despite failures? & Uptime; offline duration; override tests; patch latency & Anomaly detection; fault detection; fail-safe recs & Compromised systems damage crops & Use NIST CSF, ISA/IEC 62443 \\
Economic viability & Can facility sustain without speculative pricing? & Capex/kg; opex/kg; gross margin; anchor contracts; DSCR & Labor support; demand forecasting; waste reduction & Technically capable facility fails as business & Require crop-specific economics \\
Governance \& deployment & Is facility embedded in resilience planning? & Institutional contracts; training; audit cadence; insurance & Decision support; reporting dashboards & Isolated private asset with no public function & Define roles for all stakeholders \\
\bottomrule
\end{tabular}
\end{table*}

This framework changes the evaluation question. A conventional CEA paper may ask whether AI improves yield, climate stability, or energy use. CEA-RIF asks whether those improvements create infrastructure value under defined risks. For example, reducing VPD excursions is technically useful, but it becomes a resilience metric only when linked to yield stability, crop quality, recovery time, or reduced operator intervention during disruption. Similarly, demand response is not merely an energy-cost tactic; it becomes infrastructure-relevant when the facility can reduce or shift load without damaging crops and can document the operating envelope under which that flexibility is safe.

CEA-RIF~2.0 is also a stress test. A facility can score well on climate isolation while failing on energy cost, cyber recovery, or asset liquidity. Another facility may be commercially viable under normal retail channels but irrelevant to regional continuity if it has no institutional customers or emergency distribution plan. The framework therefore separates four claims that are often blended together: technical controllability, commercial viability, environmental performance, and public resilience value. A credible resilience asset must make all four claims explicit and measurable.

\begin{figure}[!t]
\centering
\includegraphics[width=\columnwidth]{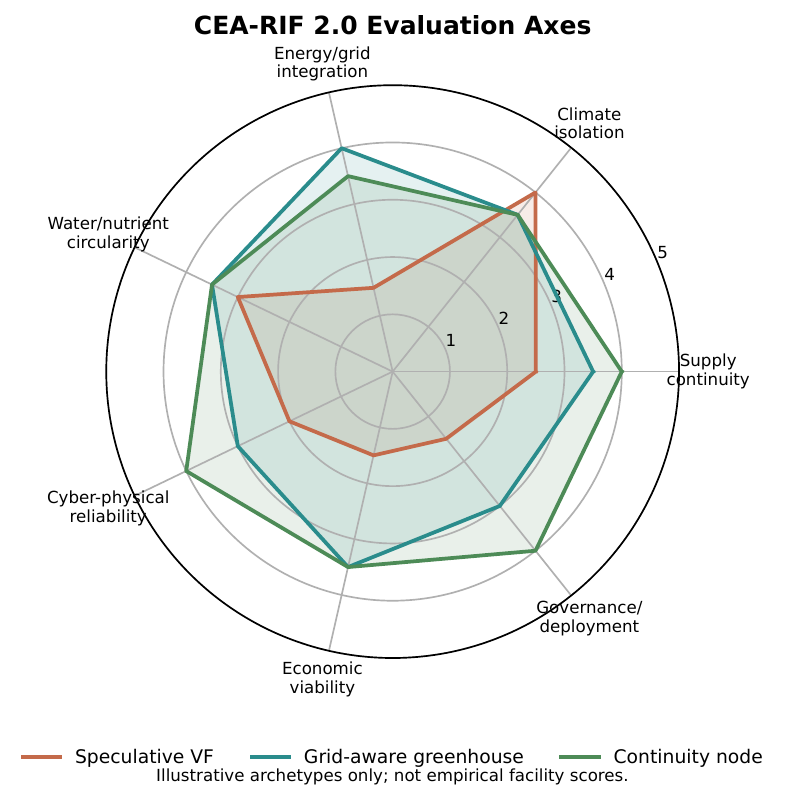}
\caption{CEA-RIF~2.0 radar chart comparing three archetypes: speculative venture vertical farm, grid-aware greenhouse, and regional continuity node.}
\label{fig:cea_rif_radar}
\end{figure}

Fig.~\ref{fig:cea_rif_radar} shows a comparative radar visualization.

% ===========================================================================
% SECTION 7: AI FUNCTIONS THAT MATTER
% ===========================================================================
\section{AI Functions That Matter}

AI should be defined by function, not by promotional language. In resilient CEA, AI is useful when it improves measured operational outcomes. Algorithmic novelty alone is insufficient. Table~\ref{tab:ai_functions} maps eight AI functions to data requirements, infrastructure value, validation metrics, and risks.

\begin{table*}[!t]
\centering
\caption{AI Functions in Resilient CEA}
\label{tab:ai_functions}
\footnotesize
\begin{tabular}{@{}p{2.8cm}p{2.8cm}p{2.8cm}p{3cm}p{3cm}@{}}
\toprule
AI Function & Data Required & Infrastructure Value & Validation Metric & Risk \\
\midrule
Predictive climate control & $T$, RH, VPD, CO$_2$, light, airflow, weather, crop stage & Reduces climate excursions and operator burden & Forecast error; excursion reduction & Poor model transfer across facilities \\
Model predictive control & Crop model, equipment constraints, climate state, energy prices, weather & Coordinates plant health with energy/operational constraints & Setpoint tracking; yield/quality preservation; energy savings & Unsafe autonomy if constraints are wrong \\
Computer vision crop monitoring & RGB/depth images, growth-stage labels, disease/pest examples & Early intervention and labor efficiency & Precision/recall; false alarm rate; time-to-detection & Bias from limited cultivars or camera setups \\
Energy-price-aware scheduling & Electricity tariffs, grid events, equipment state, crop tolerance & Enables cost control and demand response & Cost per kg; load-shift hours; yield penalty & Shifting load outside plant-safe bounds \\
Irrigation/fertigation optimization & EC, pH, substrate moisture, runoff, recipe, crop stage & Improves water/nutrient circularity & L/kg; NUE; salinity incidents & Pathogen spread or recipe drift \\
Fault and anomaly detection & Sensor streams, actuator logs, alarms, network status & Improves uptime and recovery & MTTD; MTTR & Alert fatigue or missed rare events \\
Human-supervised decision support & Integrated production, energy, crop, and risk data & Preserves operator authority while scaling expertise & Override rate; operator acceptance; outcome tracking & Opaque recommendations reduce trust \\
Model governance & Model versions, training data, performance logs, deployment history & Supports safety, compliance, and incident response & Version traceability; rollback tests & Untracked models create hidden risk \\
\bottomrule
\end{tabular}
\end{table*}

The strongest AI contribution is not ``fully autonomous farming.'' It is a measured reduction in uncertainty and avoidable operational loss. AI can forecast climate states, detect anomalies, identify early crop stress, assist growers, schedule load, reduce manual monitoring, and support safe local control. Recent reinforcement-learning model predictive control work, including Mallick~\etal's 2025 greenhouse climate-control framework~\cite{ref_mallick}, shows that advanced controllers can learn parameterizations of constraints, prediction models, and cost functions online. For resilience purposes, that promise must be paired with strict deployment requirements: hard physical constraints, local fallback modes, human-supervised override, model rollback, sensor-integrity checks, and post-event audit logs.

% ===========================================================================
% SECTION 8: GRID-AWARE CEA
% ===========================================================================
\section{Grid-Aware CEA}

Energy is the central constraint in CEA infrastructure planning. Fully enclosed vertical farms can require substantial electricity because lighting replaces sunlight and HVAC/dehumidification must manage heat and moisture. A 2024 energy-benchmarking study by Miserocchi and Franco reported a representative range of 10--18~kWh/kg for lettuce production in vertical farms, corresponding to 850--1150~kWh/m$^2$/year, and proposed a technical benchmark of 3.1--7.4~kWh/kg under expected improvements~\cite{ref_miserocchi}. Lovat, Noor, and Milo derived from first principles a current minimum cost of approximately \$10/kg dry plant matter for vertical farming under exclusively artificial light~\cite{ref_lovat}. Mills' meta-analysis of 116 studies across 40 countries found that CEA currently provides less than 1\% of U.S.\ food crops while consuming more energy than all open-field cultivation~\cite{ref_mills}. A comparative life-cycle assessment by Gargaro~\etal\ found that even when powered by renewable electricity, vertically farmed lettuce produced approximately 0.93~kg~CO$_2$e/kg compared with 0.57~kg for field-grown lettuce, despite 20-fold higher land efficiency and 8-fold lower water use~\cite{ref_gargaro}.

Energy intensity is a weakness if unmanaged. It can also become part of the paper's original contribution if CEA is framed as a controllable load. CEA facilities contain schedulable subsystems: lights, HVAC, dehumidification, pumps, fans, CO$_2$ delivery, irrigation, chilled water, thermal mass, batteries, and backup power. Some crop responses tolerate short shifts in light or climate if the controller respects daily light integral, photoperiod, VPD, temperature, humidity, and crop-stage limits. The Applied Energy demand-response study by Penuela~\etal\ showed that indoor agriculture can participate in both implicit and explicit demand response without adversely affecting vegetative growth, reporting energy cost savings of 15.34\% under implicit DR, 19.44\% under explicit DR, and 23.03\% under combined strategies~\cite{ref_penuela}. The study estimated industry-wide savings potential of \$0.96--\$16.1~million annually, with grid-side benefits of \$5.8--\$110.9~million annually. The authors also cautioned that demand response does not automatically reduce carbon footprint.

The energy evidence requires a clear distinction between resilience and climate mitigation. CEA may strengthen supply continuity or water security while still performing poorly on carbon intensity if powered by high-emission electricity. CEA-RIF~2.0 should not mandate one universal technology. It should require site-specific energy and carbon accounting, explicit grid-stress behavior, and crop-safe operating envelopes.

Grid-aware CEA means that facility electricity demand is treated as a controllable operating variable constrained by plant physiology. It has five requirements:

\begin{enumerate}
\item \textbf{Plant-safe load flexibility}: load shifting must be bounded by crop health, yield, and quality constraints.
\item \textbf{Forecast-based control}: controllers need weather, crop-state, equipment, electricity-price, and grid-event forecasts.
\item \textbf{Energy- and carbon-aware optimization}: decisions should account for cost and carbon intensity, not only kWh.
\item \textbf{Backup and fail-safe operation}: facilities must preserve safe climate modes during outages or grid events.
\item \textbf{Verified yield and quality preservation}: demand-response participation should be measured against crop outcomes.
\end{enumerate}

Grid-aware CEA does not erase energy intensity. It asks whether some portion of that demand can be scheduled, curtailed, buffered, or supplied in ways that reduce grid stress and operating cost without damaging crops. For policymakers, the implication is that CEA subsidies or resilience programs should not reward square footage alone. They should require energy accounting, load-flexibility plans, verified crop-safe operating envelopes, and evidence of carbon-aware operation where feasible.

\begin{figure}[!t]
\centering
\includegraphics[width=\columnwidth]{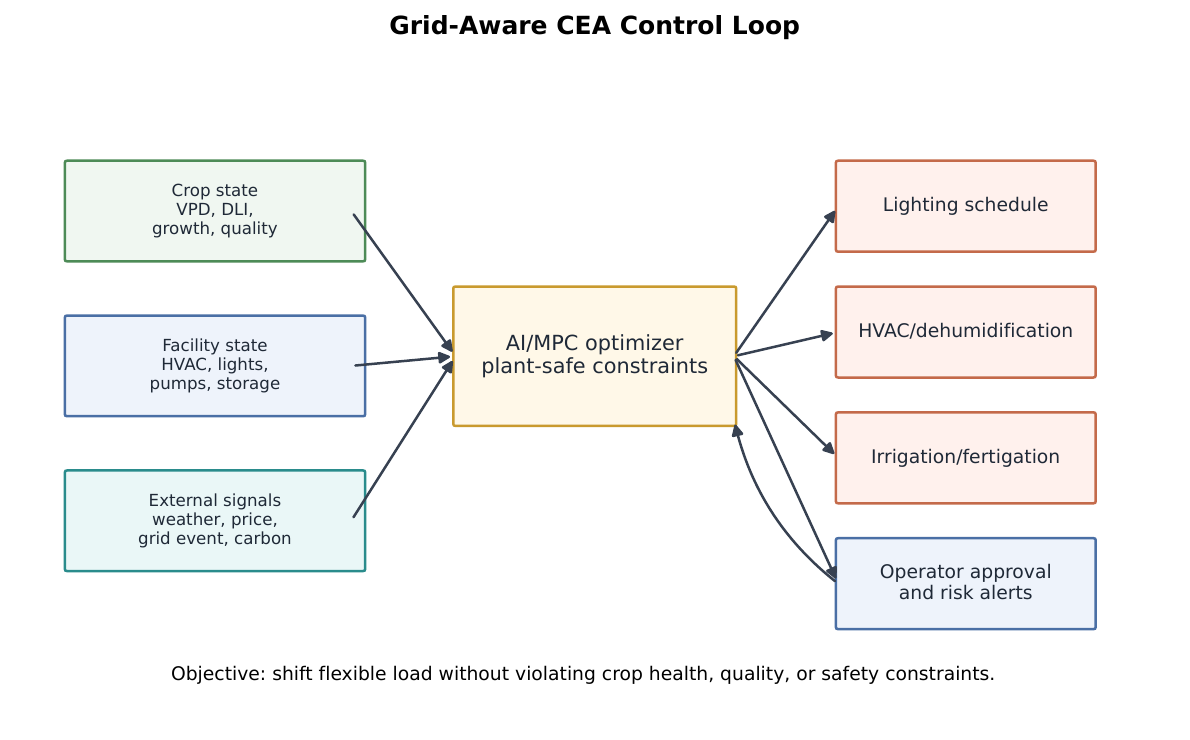}
\caption{Grid-aware CEA control loop. Inputs: crop state, climate state, weather forecast, electricity price, grid event signal, equipment constraints, carbon intensity. Controller: MPC/AI optimizer with plant-safe constraints. Outputs: lighting schedule, HVAC/dehumidification schedule, irrigation/fertigation plan, risk alert, operator approval.}
\label{fig:grid_control}
\end{figure}

Fig.~\ref{fig:grid_control} illustrates the grid-aware control loop.

% ===========================================================================
% SECTION 9: CYBER-PHYSICAL RELIABILITY
% ===========================================================================
\section{Cyber-Physical Reliability and OT Cybersecurity}

AI-driven CEA is operational technology. Modern facilities combine sensors, PLCs, greenhouse process computers, remote dashboards, cameras, irrigation and fertigation systems, lighting networks, HVAC, dehumidification, cloud services, vendor APIs, energy systems, and AI models. A cyber or data-integrity incident can become a biological production incident. Bad sensor data can trigger bad climate control. Compromised remote access can alter setpoints. Ransomware can interrupt irrigation, HVAC, lighting, or harvest operations. Model drift can reduce crop performance. Loss of connectivity can expose unsafe cloud dependence.

The threat is not theoretical. The JBS Foods ransomware attack in May~2021 shut down 13~U.S.\ meat processing plants and resulted in an \$11~million ransom payment; the NEW Cooperative attack in September~2021 compromised a grain cooperative's soil-mapping source code with a \$5.9~million demand; the Schreiber Foods attack in October~2021 shut down plants at a \$5~billion dairy company. Astral Foods reported that a March~2025 cyber incident disrupted poultry operations and affected profits by approximately R20~million~\cite{ref_astral}. FBI and CISA issued joint advisories in September~2021 and April~2022 specifically warning about ransomware targeting the food and agriculture sector. National Security Memorandum~22 elevated minimum security and resilience requirements across critical infrastructure sectors~\cite{ref_nsm22}. The bipartisan Farm and Food Cybersecurity Act of 2025~\cite{ref_farm_cyber_act} would direct USDA to conduct biennial cybersecurity threat assessments.

Cybersecurity should therefore be treated as biological production reliability, not merely IT compliance. CEA-RIF~2.0 uses NIST CSF~2.0~\cite{ref_nist_csf} for general cybersecurity risk management, CISA food and agriculture cybersecurity resources~\cite{ref_cisa_food_ag} for sector relevance, ISA/IEC~62443~\cite{ref_isa_62443} as the most relevant industrial automation and control-system standard family, and ANSI/ASABE/ASHRAE EP653~\cite{ref_asabe_ashrae} as an emerging design reference for environmental control in fully enclosed facilities.

Recommended cyber-physical controls include:

\begin{itemize}
\item OT asset inventory for sensors, controllers, actuators, gateways, cameras, servers, and remote access paths.
\item Network segmentation using zone-and-conduit thinking.
\item Secure remote access with least privilege, multi-factor authentication, and vendor access controls.
\item Sensor redundancy for critical variables such as temperature, humidity, VPD, CO$_2$, irrigation, and water quality.
\item Data integrity checks and anomaly detection for impossible or conflicting sensor values.
\item Manual override and documented fail-safe setpoints.
\item Offline fallback for local control during cloud or network outages.
\item Patch management and vulnerability tracking for controllers, gateways, operating systems, and dashboards.
\item Model versioning, rollback, and audit logs for AI-assisted decisions.
\item Incident response playbooks that include crop, food-safety, worker-safety, and business-continuity impacts.
\end{itemize}

Cyber-physical reliability can become a differentiator for serious AI-driven CEA. Many CEA papers discuss yield, sensors, or energy. Fewer integrate food-system resilience with OT cybersecurity, manual fallback, data integrity, and incident recovery. If CEA is to be treated as resilience infrastructure, this layer is mandatory.

\begin{figure}[!t]
\centering
\includegraphics[width=\columnwidth]{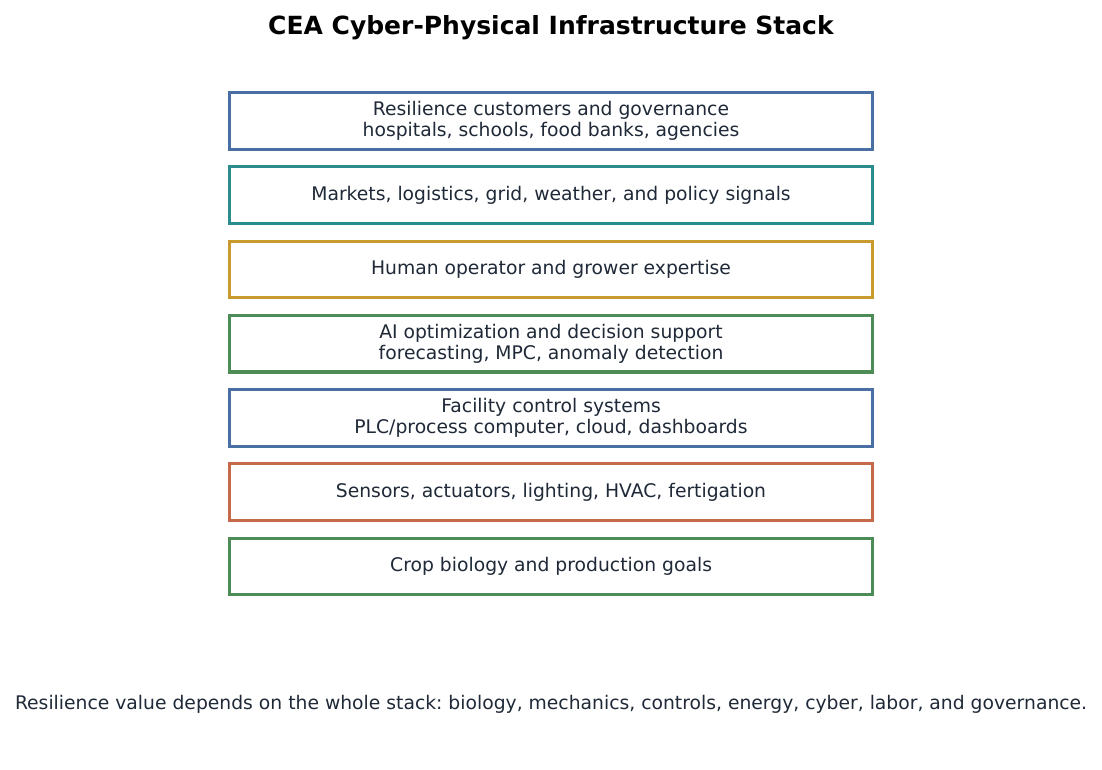}
\caption{CEA cyber-physical infrastructure stack. Layers: crop biology; sensors and actuators; facility control; AI optimization; energy and water systems; human operator; external markets, grid, weather, and logistics; resilience customers.}
\label{fig:cyber_stack}
\end{figure}

Fig.~\ref{fig:cyber_stack} shows the cyber-physical infrastructure stack.

% ===========================================================================
% SECTION 10: REGIONAL CONTINUITY NODES
% ===========================================================================
\section{Regional Fresh-Produce Continuity Nodes}

The regional fresh-produce continuity node is a deployment model for targeted CEA resilience. It is defined as a CEA facility or network designed to provide selected fresh produce, seedlings, or specialty crops during normal operations and under defined disruption scenarios while meeting standards for energy, water, cyber-physical reliability, and economic viability.

A continuity node is not necessarily a massive vertical farm in every city. It may be a high-tech greenhouse, a propagation facility, a hybrid greenhouse-indoor operation, a distributed network of smaller facilities, or a public-private cluster connected to institutional customers. Potential sites include hospitals, school districts, universities, military bases, ports and airports, disaster-prone metro areas, regional food hubs, food banks, cold-climate communities, and remote areas with fragile logistics.

The central economic idea is anchor demand. CEA economics improve when facilities have predictable buyers and crop plans. Hospitals, schools, military bases, universities, correctional institutions, food banks, grocery distribution centers, and regional food-service systems may provide more stable demand than speculative retail expansion. Public procurement alone is not a solution, and poorly designed subsidies can waste money. But anchor-demand models can help align private production with public resilience goals.

Regional continuity nodes should report metrics that matter in disruptions:

\begin{itemize}
\item kilograms of marketable produce available per week under normal and disrupted conditions;
\item days of institutional demand supported;
\item production recovery time after outage, equipment failure, or supply disruption;
\item backup power and offline control duration;
\item demand-response capability and safe load-shifting hours;
\item climate excursion frequency during stress events;
\item crop mix flexibility;
\item contract coverage and distribution reach;
\item cyber incident response time and manual override test results.
\end{itemize}

These metrics should be aligned with FEMA's Community Lifelines vocabulary~\cite{ref_fema_lifelines} without implying FEMA endorsement. FEMA's Food, Hydration, Shelter lifeline includes food, commercial food distribution, food distribution programs, commercial food supply chain, and agriculture as components or subcomponents. A CEA continuity node can therefore be evaluated against a local emergency-management scenario: for example, whether it can maintain a defined fresh-produce output for priority institutions during a 72- to 96-hour logistics disruption, while preserving safe climate control, backup power, distribution access, and reporting.

This model also creates a pathway for public-private testbeds. A university, utility, state agriculture agency, emergency-management office, grower, and CEA vendor could jointly evaluate a continuity node against CEA-RIF metrics. The result would be more useful than generic claims about local food or automation because it would define what the facility is resilient against, what crops it can realistically support, and which tradeoffs remain.

\begin{figure}[!t]
\centering
\includegraphics[width=\columnwidth]{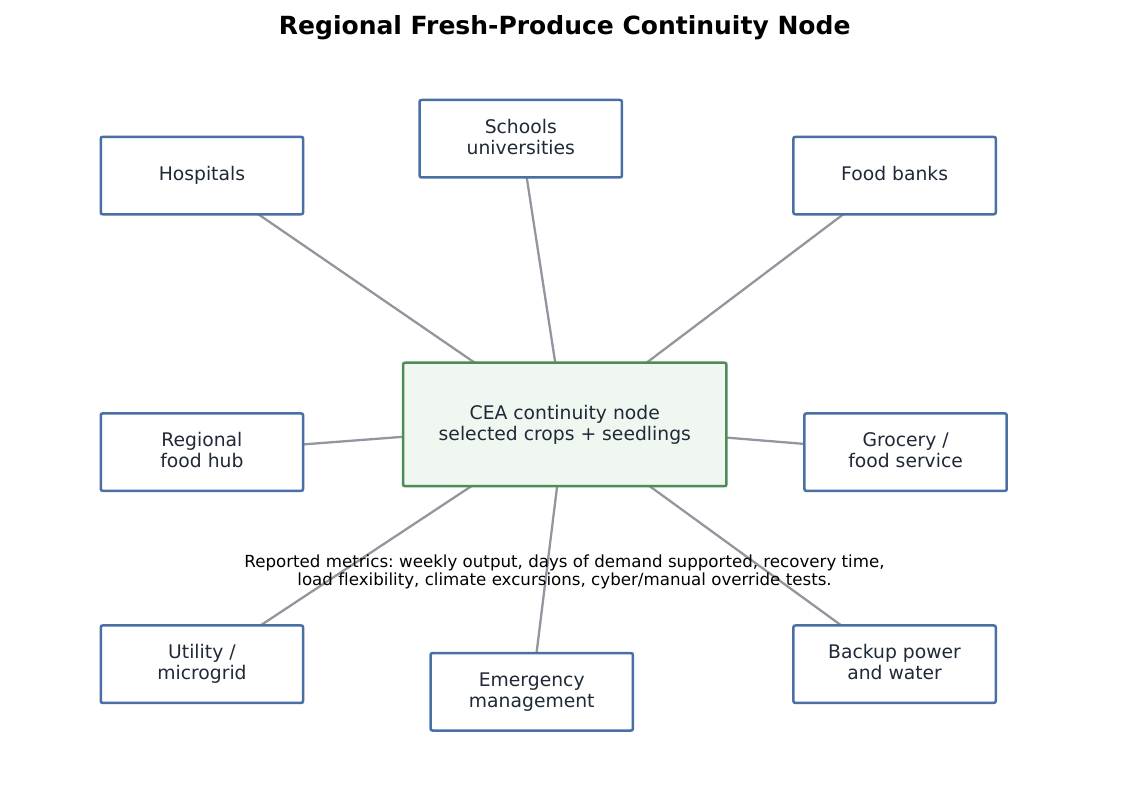}
\caption{Regional fresh-produce continuity node connected to hospitals, schools, food banks, military base, grocery/warehouse, microgrid/backup power, utility, emergency-management interface, and regional food hub.}
\label{fig:continuity_node}
\end{figure}

Fig.~\ref{fig:continuity_node} illustrates the continuity node concept.

% ===========================================================================
% SECTION 11: SUPPORTING TABLES
% ===========================================================================
\section{Supporting Tables for Evaluation and Governance}

Table~\ref{tab:pressures} links U.S.\ resilience pressures to CEA relevance, AI role, and limitations. Table~\ref{tab:failures} documents recent CEA failures and consolidation events. Table~\ref{tab:standards} maps standards and governance references.

\begin{table*}[!t]
\centering
\caption{U.S.\ Resilience Pressures and CEA Relevance}
\label{tab:pressures}
\footnotesize
\begin{tabular}{@{}p{2.2cm}p{3.5cm}p{3cm}p{3cm}p{3.5cm}@{}}
\toprule
Pressure & U.S.\ Example & CEA Relevance & AI Role & Limitation \\
\midrule
Climate extremes & Heat, drought, floods, wildfire smoke & Protected production reduces exposure for selected crops & Predictive control, stress detection & CEA remains exposed to energy/facility risks \\
Import exposure & Fresh fruit 59\%, veg 35\% import share (2023) & Regional production can buffer selected perishables & Crop planning, demand forecasting & Imports also support availability \\
Regional concentration & 90\% lettuce from two regions & Distributed CEA adds redundancy & Scenario planning, scheduling & Requires crop-specific market analysis \\
Biosecurity & HPAI: 179M+ birds; \textit{E.~coli} outbreaks & Enclosed production reduces some exposure pathways & Disease-risk alerts, sanitation logs & CEA can still spread pathogens \\
Labor constraints & Skilled grower and farm labor shortages & Automation augments scarce expertise & Decision support, monitoring, robotics & AI cannot replace agronomic skill \\
Cold-chain disruption & Port, highway, border, fuel failures & Local continuity nodes shorten chains & Inventory and distribution analytics & CEA output volume may be modest \\
Disaster response & FEMA Food, Hydration, Shelter lifeline & CEA may support institutional fresh-food continuity & Scenario dashboards & FEMA does not designate CEA as lifeline asset \\
Cyber-physical risk & Food/ag sector ransomware escalation & CEA must treat cybersecurity as production reliability & Anomaly detection, model governance & More automation increases attack surface \\
Credit constraints & USDA paused CEA loan guarantees (40\% delinquency) & Resilience projects need stronger underwriting & Financial forecasting & Public finance can amplify losses \\
\bottomrule
\end{tabular}
\end{table*}

\begin{table*}[!t]
\centering
\caption{Recent CEA and Vertical-Farming Failures and Consolidation Lessons (2022--2026)}
\label{tab:failures}
\footnotesize
\begin{tabular}{@{}p{2.2cm}cp{4.2cm}p{4cm}p{3.8cm}@{}}
\toprule
Company & Year & Event & Sector Lesson & Framework Implication \\
\midrule
Fifth Season & 2022 & Shut down Oct 2022; filed Chapter 11 Nov 2023 & Robotic-first vertical farming can fail before commercial scale & Automation must be validated against operational economics \\
Kalera & 2023 & Chapter 11 Apr 2023; assets acquired by Sandton, then sold to 80 Acres (2025) & Capital-intensive VF faces liquidity risk; assets may retain value for consolidators & Economic viability must be explicit \\
AppHarvest & 2023 & Chapter 11 Jul 2023; facilities liquidated & Large high-tech facilities fail despite \$475M SPAC funding & Evaluate capex, opex, yield, labor, contracts \\
AeroFarms & 2023--25 & Chapter 11 Jun 2023; emerged Sep 2023; WARN Dec 2025 & Emergence does not guarantee long-term viability & Resilience evaluation must be ongoing \\
Upward Farms & 2023 & Ceased Apr 2023; \$141.7M raised; no bankruptcy & Aquaponic complexity overwhelms even well-funded ventures & CEA subsectors have different risk profiles \\
Bowery Farming & 2024 & Ceased Nov 2024; \$700M raised, \$2.3B peak valuation & Venture-scale leafy-greens models may fail & Do not assume premium pricing \\
Smallhold & 2024 & Chapter 11 Feb 2024; emerged under PE control & Specialty CEA also faces restructuring risk & Crop-specific economics must be evaluated \\
Plenty & 2025 & Prepackaged Chapter 11 Mar 2025; narrowed to strawberries & Even \$940M requires balance-sheet repair & Post-hype discipline is mandatory \\
Infarm & 2023--25 & Multi-jurisdiction insolvency; liquidation Feb 2025 & Asset-acquisition rescue entities can also fail & Repeat failure is a warning sign \\
Vertical Future & 2025--26 & Revenue collapsed; administration Aug 2025; liquidation Jan 2026 & Technology-platform companies fail as rapidly as production companies & Technology and production economics must both be viable \\
80 Acres / Soli & 2025 & Acquired Kalera assets; merged with Soli Organic & Experienced operators may consolidate underused assets & Asset reuse $\neq$ original project success \\
Local Bounti & 2025--26 & \$48.4M revenue, 27\% growth; ongoing risks disclosed & Revenue growth does not remove capital/profitability risk & Pair positive signals with margin analysis \\
\bottomrule
\end{tabular}
\end{table*}

\begin{table*}[!t]
\centering
\caption{Standards and Governance Map}
\label{tab:standards}
\footnotesize
\begin{tabular}{@{}p{2.8cm}p{4cm}p{4.5cm}p{4.5cm}@{}}
\toprule
Domain & Source or Standard & CEA Relevance & Required Action \\
\midrule
Critical infrastructure & FDA/CISA Food \& Agriculture Sector & Establishes Food \& Agriculture as critical infrastructure sector & Frame CEA as modality within sector \\
Federal CEA research & USDA/DOE indoor agriculture workshop; PNAS Nexus interagency article & Systems engineering, energy-water-nutrient nexus, workforce & Use as research-agenda anchor \\
Energy benchmarking & Miserocchi \& Franco; Applied Energy DR paper & Energy intensity, load flexibility, demand-response framing & Verify boundaries and units \\
Cybersecurity governance & NIST CSF 2.0 & General cybersecurity risk-management framework & Map to CEA OT/IT governance \\
Critical-infra risk & NSM-22 and DHS strategic guidance & Elevated minimum security/resilience requirements & Use as governance context \\
Food/ag cyber resources & CISA Food \& Agriculture checklist & Sector-specific bridge between food/ag and cyber risk & Pair with NIST and ISA/IEC 62443 \\
OT security & ISA/IEC 62443 & Industrial automation and control-system security & Use zones, conduits, security levels \\
Emergency framing & FEMA Community Lifelines & Food, hydration, shelter emergency vocabulary & Say CEA may support lifeline objectives \\
CEA HVAC standard & ANSI/ASABE/ASHRAE EP653 (Oct 2021) & First HVAC standard for indoor plant environments & Use as design reference \\
CEA benchmarking & RII CEA Energy \& Water Report (Aug 2023) & First aggregate measured performance benchmarks & Reference for baseline metrics \\
CEA crop insurance & USDA RMA Controlled Environment Pilot (2026) & Expanded to 48 counties, 85\% coverage, disease coverage & Shows growing federal program legibility \\
CEA credit risk & USDA Rural Development RBCS pause & 40\% delinquency for CEA projects & Evidence for stricter underwriting \\
Food/ag cyber legislation & Farm \& Food Cybersecurity Act 2025 & Biennial assessments, annual crisis exercises & Pair with NIST CSF and CISA \\
National security & USDA Farm Security Action Plan; DARPA-USDA MOU & ``Farm security is national security'' & Frame CEA as supporting national security \\
CEA geography & NREL/PO-6A20-82104 & Distributed CEA: $-$53\% water but concentrates energy & Use for WEF nexus analysis \\
Open data & WUR/4TU Autonomous Greenhouse Challenge & Enables reproducible metric demonstration & Inspect license and feasibility \\
\bottomrule
\end{tabular}
\end{table*}

\begin{figure}[!t]
\centering
\includegraphics[width=\columnwidth]{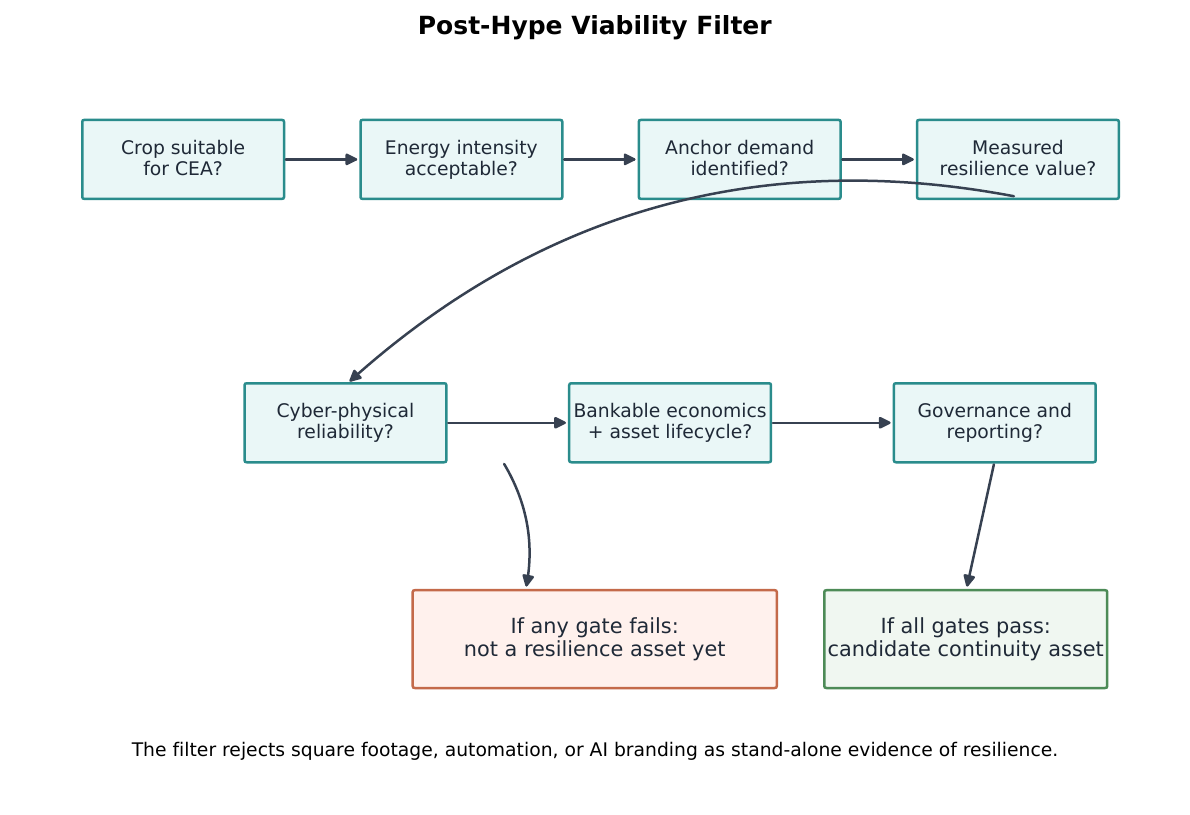}
\caption{Post-hype viability filter. Decision tree: Is the crop suitable? Is energy intensity acceptable? Is there anchor demand? Is cyber-physical reliability adequate? Does the facility provide regional resilience value? Can performance be measured? If no at any stage, the facility is not a resilience asset even if technologically advanced.}
\label{fig:viability_filter}
\end{figure}

% ===========================================================================
% SECTION 13: DISCUSSION AND LIMITATIONS
% ===========================================================================
\section{Discussion and Limitations}

CEA is neither a miracle nor a failure. It is a controlled, capital-intensive, cyber-physical production option whose resilience value depends on where, what, and how it is deployed. This framing has several implications.

First, policy should avoid blanket support for square footage, automation, or vertical-farming branding. Public support should be tied to measured outcomes: energy performance, regional continuity value, crop suitability, anchor demand, water and nutrient management, cyber-physical reliability, workforce development, and public accountability.

Second, AI should be evaluated by operational outcomes. A model that improves a benchmark but fails under sensor drift, actuator faults, crop-stage variation, or grower override is not infrastructure-grade. A simpler decision-support system that reduces climate excursions, saves operator time, documents interventions, and preserves safe fallback may be more valuable than a more novel autonomous algorithm.

Third, energy must remain central. CEA cannot credibly claim sustainability or resilience without transparent energy and carbon accounting. Grid-aware operation, demand response, thermal buffering, lighting control, HVAC optimization, backup power, and carbon-intensity-aware scheduling should become standard research questions.

Fourth, cybersecurity is part of food production reliability. As CEA becomes more connected, cyber incidents can affect crop loss, food safety, worker safety, business continuity, and regional supply continuity. Cybersecurity should be integrated into design, not added after deployment.

Fifth, regional continuity nodes need governance. A private facility can contribute to resilience, but public resilience value requires defined disruption scenarios, institutional customers, reporting standards, coordination with utilities and emergency planners, and realistic economics. Otherwise, ``local resilient food'' remains an aspirational label rather than a measurable capability.

Sixth, economic viability requires continuous scrutiny. The USDA loan-guarantee pause~\cite{ref_usda_rd_pause}, Lovat~\etal's first-principles cost floor~\cite{ref_lovat}, and Gargaro~\etal's LCA findings~\cite{ref_gargaro} require that any resilience claim be accompanied by transparent economic and environmental accounting.

The 2025--2026 evidence therefore strengthens, rather than weakens, the paper's core thesis. Bankruptcies show that technological sophistication does not substitute for crop-market fit. The USDA loan-guarantee pause shows that public finance needs better underwriting. The RMA crop-insurance expansion shows that CEA is becoming more standardized and insurable. Energy benchmarking shows that resilience value and carbon performance must be evaluated separately. Together, these signals justify an infrastructure-grade framework without justifying hype.

\textbf{Limitations.} This paper is primarily a framework paper, not a completed empirical study. The failure table is supported by official releases, court filings, and high-quality journalism but does not claim to identify definitive failure causes. International examples are policy signals, not direct proof. Energy benchmarks are crop- and system-specific. The open-data demonstration uses illustrative VPD thresholds and should be refined with crop-stage-specific thresholds. Any de-identified deployment data should be used only with permission and appropriate anonymization.

% ===========================================================================
% SECTION 14: RESEARCH AGENDA
% ===========================================================================
\section{Research Agenda}

The next research phase should prioritize eight workstreams.

\textbf{1. Interagency CEA resilience testbeds.} USDA, DOE, NASA, NSF, NIST, CISA, FEMA, state agriculture agencies, universities, utilities, and growers should collaborate on testbeds that evaluate CEA as cyber-physical food infrastructure.

\textbf{2. Open metrics and datasets.} Researchers need standardized metrics for climate stability, VPD excursions, energy intensity, water use, nutrient circularity, crop quality, model drift, manual intervention, uptime, recovery time, and demand-response performance.

\textbf{3. Grid-aware CEA pilots.} Utilities and growers should test crop-safe load flexibility, including lighting schedule shifts, pre-cooling, thermal buffering, battery coordination, and carbon-aware operation. Pilots should measure crop outcomes, not only energy cost.

\textbf{4. Cybersecurity reference architectures.} The sector needs practical OT/IT security patterns for greenhouses and indoor farms. Recent work by Bustamante~\etal~\cite{ref_bustamante} proposes the QUILLAQUA framework, and Campoverde-Molina and Lujan-Mora~\cite{ref_campoverde} provide the most comprehensive systematic literature review of agricultural cybersecurity to date.

\textbf{5. Crop-specific economic benchmarks.} CEA-RIF requires crop-specific economics. Leafy greens, microgreens, tomatoes, cucumbers, berries, herbs, seedlings, and propagation material have different margins, energy needs, labor needs, distribution channels, and resilience value.

\textbf{6. Public-private anchor-demand models.} Hospitals, schools, universities, food banks, military bases, and regional food hubs could participate in carefully designed continuity contracts with performance requirements.

\textbf{7. Water-energy-food nexus integration.} Zou~\etal~\cite{ref_zou} demonstrate next-generation water-saving strategies using a nexus approach with four emerging technologies. CEA resilience evaluation should integrate water, energy, and food dimensions rather than treating them as separate optimization targets.

\textbf{8. Workforce and grower decision support.} AI should augment growers and technicians. Research should focus on human-supervised autonomy, explainable recommendations, training, and the preservation of horticultural expertise.

% ===========================================================================
% SECTION 15: DATA AND CODE AVAILABILITY
% ===========================================================================
\section*{Data and Code Availability}

The demonstrative open-data analysis uses the Autonomous Greenhouse Challenge, Second Edition (2019) dataset published by Hemming, de Zwart, Elings, Petropoulou, and Righini through 4TU.ResearchData under a CC0 license, DOI 10.4121/uuid:88d22c60-21b3-4ea8-90db-20249a5be2a7~\cite{ref_wur_4tu_2nd}. Reproducibility materials for the illustrative VPD and resource-use table are provided as ancillary files, including the script \texttt{cea\_rif\_wur\_4tu\_metrics.py} and generated CSV/Markdown outputs. The analysis is intended as a transparent methods demonstration, not as a final agronomic ranking of greenhouse compartments.

% ===========================================================================
% SECTION 16: AI ASSISTANCE DISCLOSURE
% ===========================================================================
\section*{AI Assistance Disclosure}

Generative AI tools were used for drafting support, source-organization assistance, checklist generation, and manuscript-editing support. The author is responsible for reviewing all claims, citations, analysis, and conclusions before public submission.

% ===========================================================================
% SECTION 17: CONCLUSION
% ===========================================================================
\section{Conclusion}

AI-driven CEA should not be framed as the future of all food or as a replacement for conventional agriculture. That argument is too broad and too easy to defeat. The stronger argument is that AI-driven CEA can become a measurable, energy-aware, cyber-physical resilience layer for selected fresh-produce and specialty-crop systems when evaluated through strict infrastructure, economic, energy, cyber, and governance criteria.

CEA-RIF~2.0 provides a way to make that evaluation concrete. It asks whether a system improves supply continuity, climate isolation, grid integration, water and nutrient circularity, cyber-physical reliability, economic viability, and governance. It also forces the field to confront recent failures honestly. Post-hype CEA does not need louder claims. It needs better metrics, better controls, better energy integration, better cybersecurity, better business discipline, and better regional planning.

If those conditions are met, AI-driven CEA can support U.S.\ fresh-produce resilience in targeted settings: continuity nodes near institutional demand, greenhouses that interact intelligently with the grid, propagation systems that support post-disaster recovery, and cyber-resilient production networks that preserve safe operation under disruption. The national-interest value lies not in ``AI farms will save agriculture,'' but in disciplined infrastructure-grade systems that help selected fresh-produce and specialty-crop supply chains remain productive, secure, and recoverable under stress.

% ===========================================================================
% REFERENCES
% ===========================================================================
\bibliographystyle{IEEEtran}
\bibliography{cea-rif}

\end{document}